\documentclass{article}
\usepackage{amsfonts}
\usepackage{amsmath}
\usepackage{amssymb}
\usepackage[dvips]{color,graphicx}

\newcommand{\M}{m\displaystyle}
\newcommand{\V}{v}
\newcommand{\A}{\fontshape{sl}\textrm{a}\displaystyle}

\newcommand{\be}{\begin{equation}}
\newcommand{\ee}{\end{equation}}

\newcommand{\ba}{\begin{equation}\begin{array}}
\newcommand{\ea}{\end{array}\end{equation}}

\newcommand{\bea}{\begin{eqnarray}}
\newcommand{\eea}{\end{eqnarray}}

\newcommand{\bse}{\begin{subequations}}
\newcommand{\ese}{\end{subequations}}

\newcommand{\rcn}{\renewcommand{\arraystretch}{1.3}}
\newcommand{\rcd}{\renewcommand{\arraystretch}{2}}
\newcommand{\ds}{\displaystyle}

\renewcommand{\arraystretch}{1.2}

\title{So you want to stop time}
\author{A J Janca\\\footnotesize{Department of Physics, North Carolina State University\\Raleigh NC 27695, United States}}
\date{16 January 2007\\ \normalsize\textit{Revised} (v2) \textit{26 April 2007}}

\makeatletter
\renewcommand{\maketitle}{%
  \begin{flushleft}%
    {\huge\bfseries\@title\par}%
    \bigskip \bigskip \bigskip \bigskip \bigskip \bigskip
    {\large\@author\par}%
    \medskip
    {\large\@date\par}%
    \bigskip
  \end{flushleft}%
}
\makeatother

\begin{document}
\maketitle

\begin{small}

\noindent A model of a stasis chamber, slowing the passage of time in its interior down to arbitrarily small rates relative to the outside world, is considered within classical general relativity.  Since the model is adapted from the Majumdar-Papapetrou spacetime, interior spatial volumes are increased by the same factor (cubed) as the rate of time is reduced.  An interesting side effect is that real and apparent (gravitational) forces as perceived by an interior observer are altered, but in opposite ways.  Comparison with special-relativistic time dilation shows the use of such a static stasis chamber to be economical only when the most drastic slowing of time relative to the outside world is desired ($d\tau/dt \lesssim 10^{-20}$) or when one wants to avoid spending the time needed to accelerate to relativistic speeds. \\

\end{small}

\section{Relative dimensions}

A few years ago, a Russian engineer claimed to have invented a device that, by manipulating the ether, slowed the rate of time within a spherical chamber relative to the outside by 3\%. \cite{Chernobrov2001}  This result may seem incredible; but there is nothing wrong in principle with trying to change the local rate of time from its everyday value, and a similar effect can in fact be easily modelled within orthodox physics.  Since classical general relativity is the proper science of space and time, it is the logical place to start when looking for methods to alter their properties in some way.

For instance, close to the Schwarzschild radius of a massive gravitating body, time slows down and spatial volumes are expanded relative to points far away.  In principle, a massive spherical shell could be built to take advantage of this phenomenon: inside, spacetime would be flat, but time would run much slower than at points reasonably distant from it.  But the shell would have be either very large or be composed of some unique form of matter strong enough to resist complete gravitational collapse.

Suppose one wanted to slow time down to an arbitrarily slow rate within some enclosed volume, so slow that it could not be distinguished from a timeless state\textemdash in other words, build a stasis chamber.  Further, rule out chambers built on astronomical scales, and construction materials with internal stresses greater than those of neutron stars.  Assuming the builder restricted herself to conventionally agreed physics, what would it take?

\begin{figure}[t]
\centering \includegraphics[angle=0, width=\textwidth]{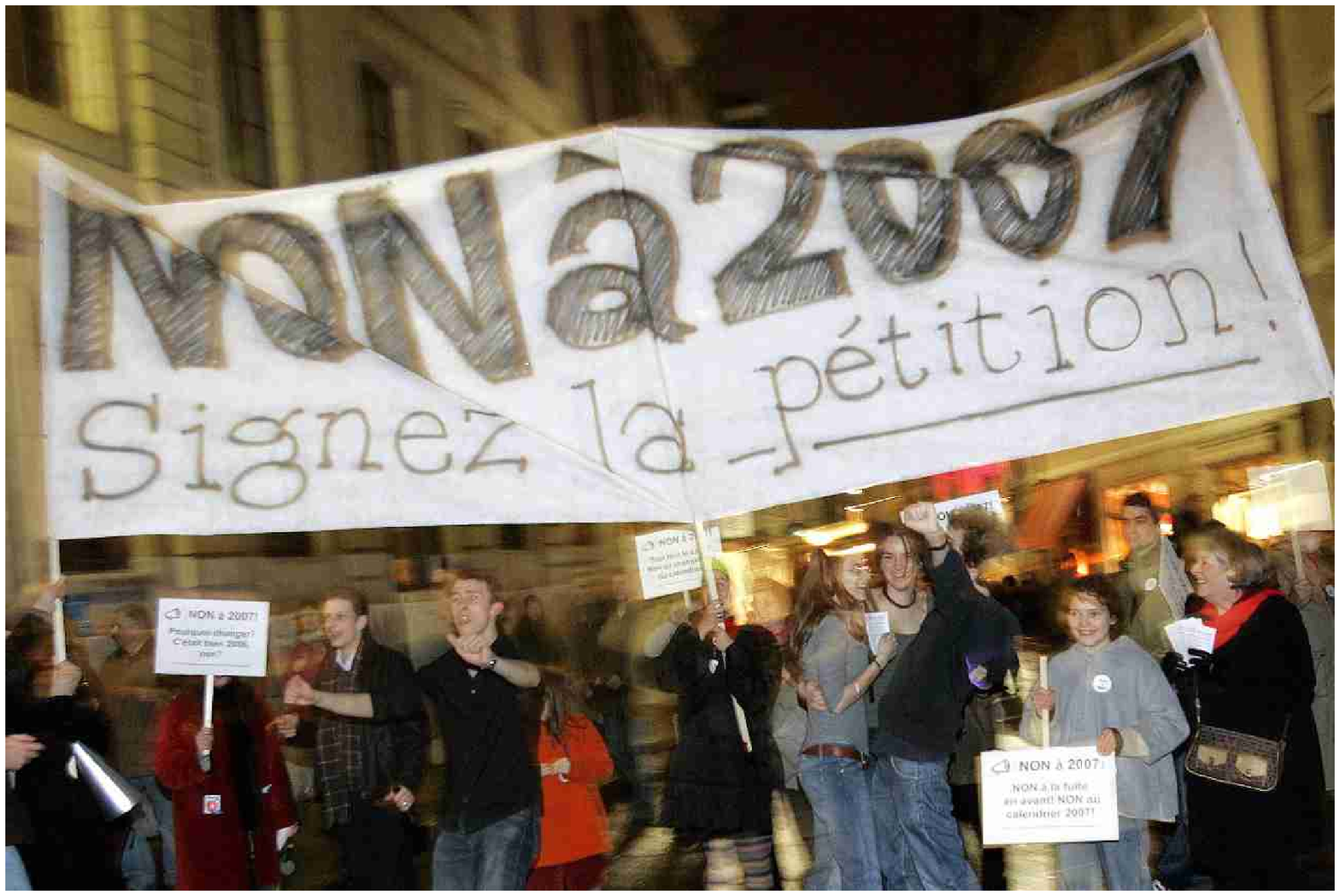}
\label{fig:Non}
\end{figure}

\section{Model}

A convenient model to work with is the Majumdar-Papapetrou (MP) or ``conformastatic'' line element \cite{Majumdar1947}, \cite{Papapetrou1947}, \cite{Synge1960}
\bea
ds^2 = a^{-2} dt^2 - a^2 (dx^2+dy^2+dz^2)
\eea
\noindent where $a(x^i)$, approaching the asymptotic value $a \rightarrow 1$ at spatial infinity, is a function of the spatial coordinates only and the coordinate speed of light in Minkowski space $c$ is set equal to 1.\footnote{Except where explicitly stated, the gravitational and electric force constants $G$ and $k_e$ will also equal 1.}  Its energy tensor, in the orthonormal frame diagonalizing it, has the simple form
\begin{displaymath}
{T}^{\hat{a}\hat{b}}=
 \left|\begin{array}{crcc}
  A + B & 0 & 0 & 0 \\
  0 & -A & 0 & 0 \\
  0 & 0 & A & 0 \\
  0   & 0   & 0 & A
\end{array} \right|
\end{displaymath}

\noindent where \vspace{-24pt}

\bse \bea
8 \pi A &=& \!\!\phantom{-}a^{-4} |\nabla a|^2 \\
8 \pi B  &=& \!\!-2\, a^{-3} \nabla^2 a
\eea \ese
Where $\nabla^2 a = 0$, this represents a static electric or magnetic field.  Where $\nabla^2 a \neq 0$ critically charged matter is also present, with mass and charge density both equalling $-(4\pi a^3)^{-1} \nabla^2 a$.  Since the gravitational and electric forces on them are equal and opposite, these physical sources are static and stable, without internal pressures or tensions.  In both cases the electric or magnetic scalar potential is $\Phi = a^{-1}$.

As can be seen from the metric, the effects these fields have on space and time are inversely proportional.  When $a<1$, spatial volumes are decreased and time speeds up relative to distant points; when $a>1$, the opposite effects occur.

So long as $\nabla^2 a \leq 0$, the relative scales of space and time may be changed in local regions without violating the energy conditions of classical general relativity.  By truncating the exterior fields at the inner wall of a chamber enclosing a volume of flat spacetime where $a$ has a constant value $a_0 > 1$, the rate of time inside relative to sufficiently distant points outside may be reduced to arbitrarily slow levels $dt_{in}/dt_{out} = a_0^{-1}$.

\subsection{Interior solution}

\begin{figure}[t]
\centering \includegraphics[angle=0, height=16pc]{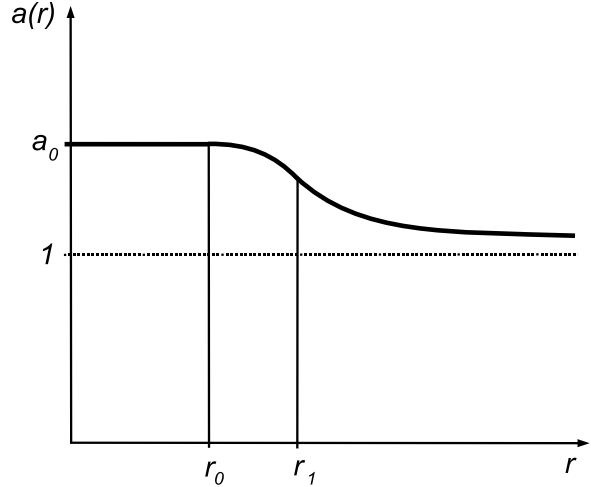}
\caption{$a(r)$ for $a_0 > 1$.  Gradient of exterior $1/r$ potential exaggerated for clarity.}  
\label{fig:a(r)}
\end{figure}

MP fields can have any spatial symmetry, but for simplicity a spherical shell will be used here.

Choose for $a(r)$ a function (figure \ref{fig:a(r)}) smoothly joining a spherically symmetric electro-vacuum spacetime outside some boundary $r=r_1$, to a flat, hollow interior space of coordinate radius $r_0 < r_1$ and $a(r)_{in} = a_0 =$ constant, by means of a spherical shell of finite coordinate thickness $\Delta r = r_1-r_0$:
\be
a(r) = 
\begin{cases}
 \quad 1+ q/r & \quad \phantom{0} r_1 < r < \infty \\
 \quad a_0 - C(r-r_0)^2 & \quad \phantom{0} r_0 \leq r \leq r_1 \\
 \quad a_0 & \quad \phantom{r_1}\hspace{-2pt} 0 \hspace{2pt} \leq r  < r_0 \\
\end{cases} 
\ee
\rcd\ba{lcc}
C &=& \ds\frac{a_0 - 1}{(r_1-r_0)(3 r_1-r_0)} \nonumber \\
q &=& \ds\frac{2 (a_0-1) \: r_1^2}{3 r_1-r_0} \nonumber
\ea\rcn
The integrated total mass of the shell is then just $M_{\scriptscriptstyle{TOT}} = q$, equal to its ADM mass and total charge.\footnote{This is generically true of all spherical MP shells matching to a flat interior.}

\section{Effect on accelerations}
\subsection{Static chamber on Earth's surface} \label{staticacceleration}

Among the consequences of altering local length and time scales with an MP field is that an observer within the shell, even though space is flat there, will perceive external forces differently than will an observer far away. \cite{Ivanov2004}, \cite{Ivanov2005}  But it is surprising that apparent forces due to spacetime curvature\textemdash grav\-i\-ta\-tion\-al forces\textemdash are affected differently than forces due to real acceleration.

MP fields are linear, as in Newtonian gravity and classical electrostatics, and hence are superimposable.  This allows the Earth's weak surface gravity to be approximated by a uniform MP field $a_{\scriptscriptstyle{E}}(z) = g z$, such that $a$ in the empty interior will be the simple sum of both fields $a_{in}(x^i)=a_0 + a_{\scriptscriptstyle{E}}(z) = a_0 + g z$.  The gravitational force felt by a static observer of mass $\M$ inside the chamber (in her own local frame, where her four-velocity and acceleration are $u^{\hat{a}}=u^{\hat{t}}=1$ and $d^2 x^{\hat{a}}/ d\tau^2 = 0$) will be given by
\bea
F^{\hat{z}} &=& \M \left[\frac{d^2 \hat{z}}{d\tau^2} + \Gamma^{\hat{z}}_{\phantom{z} \hat{t} \hat{t}} (u^{\hat{t}})^2 \right] \nonumber \\
&=& a^{-2}_{in} \M g \;\; \approx \;\; a_0^{-2} \M g
\eea
Since the magnitude of Earth's surface gravitational acceleration $g$ in geometric units is on the order of $10^{-16}$, any $a_0$ differing substantially from 1 will dwarf $gz$ so that $a_{in} \approx a_0$.  Thus for $a_0 > 1$, the Earth's gravitational field on an inside observer is effectively dampened by a factor of $(a_0)^2$.  

\subsection{Accelerative gee-forces}

Now remove the chamber from the surface of the Earth, and install it inside an accelerating spaceship.  For a ship moving at $\V=dx/dt \ll c$ and accelerating at $\A=d^2 x/dt^2$ in free space with vanishingly weak external gravitational fields, the accelerative force felt by an observer within will be
\bea
F^{\hat{x}} &=& \M \frac{d^2 x^{\hat{a}}}{d\tau^2} \nonumber \\
&=& \M \left(\frac{d\hat{x}}{dx}\right)^2 \left(\frac{dt}{d\tau}\right)^2 \frac{d^2 x}{dt^2} \nonumber \\
&=& (a_0)^4 \, \M \A 
\eea
\noindent Hence for $a_0 > 1$, in contrast to the situation with static gravitational forces, the effects of accelerative gee-forces on the internal observer are \emph{increased}.  The astronaut hoping to dampen these to make spaceflight at high acceleration more comfortable (or at least survivable) would have to build her `acceleration couch' out of unphysical negative-mass matter (section \ref{refinements}).

\section{Comparative energy efficiency}

\subsection{Relativistic spaceflight}

To get a sense of its practicality as a method of slowing time, one can compare the energy requirements of a human-sized stasis chamber (how much mass would be needed to build one) with those of special-relativistic time dilation (how much mass must be consumed to reach relativistic speeds).  

Consider an astronaut who accelerates at one Earth gravity to a speed sufficiently close to that of light that her personal rate of time compared to that back on Earth, $d\tau/dt = 1/u^0$, is a small fraction of 1.  To reach a time-dilational factor of $u^0 (\tau_{\scriptscriptstyle{A}}) = \cosh (g \tau_{\scriptscriptstyle{A}}) = 100$, for instance, she must accelerate for a little over five years of proper time $\tau_{\scriptscriptstyle{A}}$. \cite{MTW}  She then coasts at this speed for some desired period, before decelerating to a stop, turning around, and repeating the process to return home.  

Suppose that her ship is capable of completely efficient mass-to-energy conversion, and that she can harvest the needed reaction mass from the dust of interstellar space.  Then each leg of acceleration will require burning $\Delta m = (\gamma_f -1) \: m_0 = (u^0_f - 1) \: m_0$, where $u^0_f \equiv u^0 (\tau_{\scriptscriptstyle{A}}) = \gamma_f$ denotes the time component of the astronaut's four-velocity in the rest frame of the Earth at the completion of the acceleration and $m_0$ is the combined rest mass of astronaut and ship.  For $u^0 \gg 1$,
\bea
\Delta m_{\scriptscriptstyle{TOT}} &=& 4 \, m_0 (u^0_f - 1) \; \approx \; 4 \, m_0 \, u^0_f \label{sr}
\eea
\noindent for the entire round trip.\footnote{If the astronaut's trip consists only of the accelerations and decelerations, the ratio of total Earth time to proper time elapsed is $\Delta t_{\scriptscriptstyle{E}}/\Delta \tau = (g \tau_{\scriptscriptstyle{A}}/c)^{-1} \sinh (g \Delta \tau_{\scriptscriptstyle{A}}/c)$, where $\tau_{\scriptscriptstyle{A}}=\Delta \tau /4$ is the time of each accelerational segment.  As the maximum speed reached gets closer to $c$, this ratio becomes negligible compared with that experienced by the astronaut while she is coasting at that speed, $u^0_f = \cosh (g \tau_{\scriptscriptstyle{A}}/c)$, and so can be neglected for this rough calculation.}  Then for $u^0_f = 100$ and $m_0$ equalling, say, 100,000 kg, the total mass consumed as fuel will be $4 \cdot 10^7$ kg. \\

\subsection{Stasis chamber}

The interior radius of the static stasis chamber as measured by a person inside is $\hat{r}_0 = (g_{rr})^{1/2} r_0 = a_0 r_0$.  Choose $\Delta {r}$ to be sufficiently small so that $r_0 \approx r_1$ and $a_0 \approx 1 + q/r_0$.  Then for some given interior radius $\hat{r}_0$ and dilation factor $a_0$, the total charge and mass of the shell must be
\bea
q_{\scriptscriptstyle{TOT}} &=& (1-1/a_0) \: \hat{r}_0
\eea
Hence for any substantial dilation ($a_0 \gg 1$) the chamber's actual internal radius $\hat{r}_0$ approximately equals its total mass and charge.  From the conversion factors from relativistic units (measured in meters) to conventional units
\ba{lllll}
1 \; \mathrm{m} &=& c_0^2/G &=& 1.347 \cdot 10^{27} \; \mathrm{kg} \\
&=& \sqrt{k_e G}/c_0^2 &=& 1.161 \cdot 10^{17} \; \mathrm{C}
\ea
\noindent it can be seen that they would have to be enormous for any but the tiniest of chambers.  For instance, a sphere of internal radius $\hat{r}_0 = 5$ m would need a total mass of $q = 4.94 $ m $=6.65 \cdot 10^{27}$ kg, a thousand times the mass of the Earth, to achieve the same time dilation effect of $a_0  = 100$ as the astronaut in the example above.

Past this point, one can indeed set the \emph{relative} size of the internal space\textemdash since it scales inversely to the relative rate of time\textemdash to be as large as one likes, even though the absolute interior volume of the chamber remains fixed by $q$.  It is true that to hold any macroscopic object, $q$ itself must have macroscopic dimensions.  But once this necessary investment in $q$ has been made, the relative rate of time can be reduced to as close to zero as one likes without further substantial additions to the chamber's mass and charge.\footnote{This assertion may seem more reasonable when one considers that the shell's exterior is just the extremal ($q=m$) Reissner-Nordstr\"om spacetime in isotropic coordinates \cite{Prasanna1968}, where $r_{\scriptscriptstyle{RN}} = r_{\scriptscriptstyle{MP}} + q$.  \emph{If} continued all the way to the origin, $r_{\scriptscriptstyle{MP}}=0$ would mark the event horizon of a charged black hole, where the coordinate time $t$ used by distant observers crawls to a halt.}  

In any case, the electric and gravitational fields needed to warp spacetime to such an extent would be enormous in the vicinity of the sphere.  Since the material making up the sphere is stabilized by its balanced mass and charge, it will itself be fine, as will be anyone inside; but any nearby body will be sucked in and pancaked against its outside wall.

Under most circumstances, enclosing oneself within such a massive\footnote{Recall that the electric charge and field are included only to stabilize the shell, and to simplify the calculations in section \ref{staticacceleration}\textemdash their effects on $g_{tt}$ are minor by comparison with that due to the shell's mass.} shell would be about as sophisticated a method of slowing time as finding a black hole and standing next to it for a while.  In energy terms, it would become more efficient than special-relativistic time dilation only if one wished to slow the passage of time to less than $\approx 10^{-20}$ times its normal rate.  Otherwise, the one advantage such a stasis chamber would have over relativistic spaceflight is that one would not need to spend decades of proper time accelerating and decelerating to achieve the same time dilation.

It is appealing to think of things one could use a stasis chamber for, like a better refrigerator\textemdash just pop your leftovers into the kitchen stasis-box, and they would always be as hot and fresh as the hour they were cooked.  On the other hand, its property of expanding spatial volumes would make it the ideal storage solution for the householder with too much stuff and not enough space, since with such a box or closet she could squeeze as much as she wanted into the tightest nook.  But its strong fields and energy requirements would make such mass-consumer applications impractical at best.  For applications important enough to justify the resources necessary to build one, such as a doomsday time capsule for the world's genetic heritage \cite{GCT} or a way to send terminally ill patients (permanently) into the future for treatment, time dilation via relativistic spaceflight would generally be a safer and more economical alternative.

\subsection{Refinements} \label{refinements}

\vspace{6pt}

\noindent \emph{A door} \vspace{6pt}

\begin{figure}
\centering \includegraphics[angle=90, height=16pc]{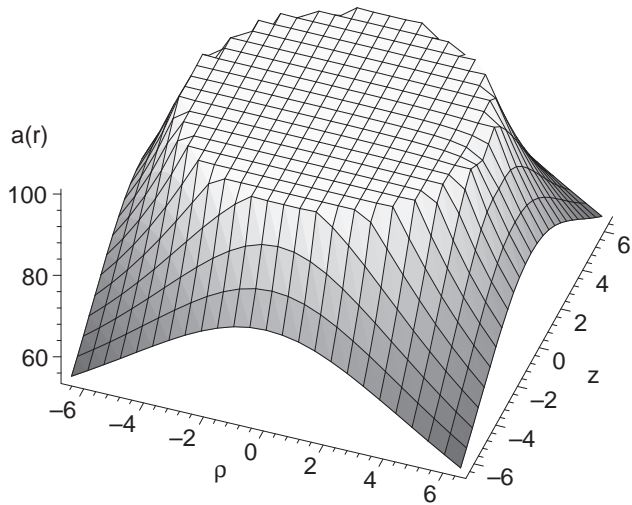}
\caption[$a(r)$, equatorial slice]{$a(r)$, equatorial slice, equipotential spherical shell of radius 5 m (mass and charge $q=4.94$ m) with a circular hole of radius 50 cm centered at $\rho=r \cos \theta = 5$ m, $z= 0$ m, visible as the small lip on the right side of the plateau.  The hole's effect on the internal potential is negligible.}  
\label{fig:Potential}
\end{figure}

\noindent Since MP spacetimes are built from solutions to Laplace's equation $\nabla^2 a=0$, it is straightforward to modify the closed-shell source to give it an entryway.  The potential problem for a charged spherical shell with a circular hole can be exactly solved \cite{PuncturedSphere} although it is easier to calculate numerically (figure \ref{fig:Potential}).  For a 5 m radius shell with a one meter diameter circular aperture, the potential $a$ hardly differs from that of a shell that is perfectly spherically symmetric\textemdash the aperture perturbs the fields of the closed body no more than do the holes in a Faraday cage\textemdash and the interior space remains almost perfectly flat. \\
\vspace{6pt}

\noindent \emph{A spherical capacitor} \vspace{6pt}

\noindent One could make this model more sophisticated by enclosing the chamber within a second, concentric shell with mass and charge opposite to those of the first, resulting in a sort of gravitational (and electric) capacitor.  Not only would this bring the total mass and charge of the system to zero, it would also confine the strong electric and gravitational fields that actually effect the change in local geometry to the space between the two walls.  The exterior space, where $a=1$, would like the interior be completely flat.    

However, the second wall would have to be made of negative-mass exotic matter, and a lot of it.  For example, assembling even a Planck-scale sphere of interior radius $\hat{r}_0 = 10^{-35}$ m would require the laboratory scientist to somehow scrape up $-0.01 \; \mu$g worth of exotic matter, a considerable amount to work with even assuming such macroscopic amounts of this peculiar substance could be created and stabilized in the first place.  The same problem would present itself for the experimenter wishing to build a chamber with the opposite effects\textemdash one with $a_0 <1$, thereby \emph{increasing} its interior rate of time relative to the outside\textemdash either by reversing the order of the shells, or by just using the negative-mass shell by itself. 

It may be that substantial modifications on the local rate of time\textemdash unlike small but \emph{measurable} distortions that do seem to be achievable in the laboratory today \cite{Ivanov2004}, \cite{Ivanov2005}\textemdash could be achieved within classical general relativity with more conservative sources.  But since in general, large distortions in local geometry require huge concentrations or total amounts of matter, this prospect seems unlikely at best.

\section*{Acknowledgements}

Although stasis chambers that reduce the flow of time inside to zero and magical rooms that are larger on the inside than they are outside are both ancient staples of fairytales and science fiction, the original idea for this note came from a physics paper on a quite different but equally exotic phenomenon, warp drive.

Recently Broeck \cite{Broeck1999} showed that one of the problems of Alcubierre's warp drive \cite{Alcubierre1994} could be fixed by a clever trick.  The problem was that a warp bubble large enough to hold a human being would require fantastic amounts of negative-mass exotic matter\textemdash more, apparently, than all the positive mass contained in the known Universe.  Broeck proposed placing a second bubble, having the special property that it was many orders of magnitude larger on the inside than it was on the outside, inside the warp bubble.  By expanding spatial volumes in this way, a large spaceship could be squeezed into a Planck-scale warp bubble needing much less exotic matter than a macroscopic one.\footnote{Similar dodges were used to reduce the exotic matter needs of static wormholes \cite{VisserKarDadhich2003} and the Krasnikov tube \cite{GravelPlante2004}, though since there is no generally accepted way to quantify the total amount of exotic matter present in the system one can't rigorously say just how much of it remains in the repaired spacetimes.\label{exotic}}  

The example presented here\textemdash somewhat modified from that of \cite{Broeck1999}, to allow time to scale in inverse proportion to space\textemdash shows that the Broeck bubble itself could be made without using exotic matter, though at the price of introducing huge charge and matter densities within its walls and correspondingly intense electric and gravitational fields outside them.

Ivanov \cite{Ivanov2004}, \cite{Ivanov2005} has written extensively on the technological reasonableness of devices exploiting MP fields, including proposed experimental tests.

I am grateful to my supervisor A Kheyfets for advice and criticism with the material of this paper, and to S V Krasnikov for pointing out the difficulty mentioned in footnote \ref{exotic}.  Photo credit Fabrice Coffrini/AFP/Getty Images.  This work was partially supported by the National Security Education Program.

%
%
%
%


\begin{thebibliography}{15}

\bibitem{Chernobrov2001}  V A Chernobrov, A V Frolov.  Cover article, New Energy Technologies 3 (Nov/Dec 2001), http://www.faraday.ru/content03.html.  Description at http://alexfrolov.narod.ru/rustmc.html (the English version referenced in v1 is no longer available online).

\bibitem{Majumdar1947}  S D Majumdar.  A class of exact solutions of Einstein's field equations.  Physical Review v 72 (1947) pp 390-398.

\bibitem{Papapetrou1947}  A Papapetrou.  A static solution of the equations of the gravitational field for an arbitrary charge-distribution.  Proceedings of the Royal Irish Academy A v 51 (1947) pp 191-204.

\bibitem{Synge1960}  J L Synge.  Relativity: the general theory.  Interscience, New York 1960.  

\bibitem{Ivanov2004}  B V Ivanov.  Strong gravitational force induced by static electromagnetic fields, arXiv:gr-qc/0407048 (2004).  

\bibitem{Ivanov2005}  B V Ivanov.  On the gravitational field induced by static electromagnetic sources, arXiv:gr-qc/0502047 (2005). 

\bibitem{MTW}  C W Misner, K S Thorne, J A Wheeler.  Gravitation.  W H Freeman, New York 1973.  

\bibitem{Prasanna1968}  A R Prasanna.  A solution for an isolated charged body in isotropic co-ordinates.  Current Science v 37 (1968) pp 430-431.

\bibitem{GCT}  Such as the new International Arctic Seed Vault currently being built by the Norwegian government on Spitzbergen island, due to be completed late 2007.  The Global Crop Diversity Trust, http:// www.croptrust.org/.  See also BBC News 19 June 2006, ``Work begins on Arctic seed vault'', http://news.bbc.co.uk/2/hi/science/nature/ 5094450.stm, and 9 February 2007, ```Doomsday' vault design unveiled'', http://news.bbc.co.uk/2/hi/science/nature/6335899.stm.
%

\bibitem{PuncturedSphere}  E W Hobson.  The theory of spheroidal and ellipsoidal harmonics.  Cambridge University Press 1931.  Section 267.  

This reference gives not the complete solution for the potential of a charged spherical bowl, but only the set of basis functions needed for a series solution in toroidal harmonics.  Unfortunately simpler alternatives in the literature are unsuitable for various reasons.  For instance, the approximate solution by superposition of spherical harmonics given in \cite{MorseFeshbach1953} does not have suitable boundary conditions for this case, since the surface defined by the shell's aperture is set at constant potential.  

\bibitem{MorseFeshbach1953}  P M Morse, H Feshbach.  Methods of theoretical physics.  McGraw-Hill, New York 1953.  Part II, section 10.3.  






\bibitem{Broeck1999} C van den Broeck.  A `warp drive' with more reasonable total energy requirements.  Classical and Quantum Gravity v 16 (1999) 3973-3979.  Also arXiv:gr-qc/9905084 v5. 

\bibitem{Alcubierre1994}  M Alcubierre.  The warp drive: hyper-fast travel within general relativity.  Classical and Quantum Gravity v 11 (1994) L73-L77.  Also arXiv:gr-qc/0009013.

\bibitem{VisserKarDadhich2003}  M Visser, S Kar, N Dadhich.  Traversable wormholes with arbitrarily small energy condition violations.  Physical Review Letters v 90 (2003) 201102.  Also arXiv:gr-qc/0301003.

\bibitem{GravelPlante2004}  P Gravel, J Plante.  Simple and double walled Krasnikov tubes.  I. Tubes with low masses.  Classical and Quantum Gravity v 21 (2004) L7-L9.


\end{thebibliography}
\end{document}